\documentclass[preprint]{ptephy_om}

\preprintnumber{RIKEN-iTHEMS-Report-25, KYUSHU-HET-308} 

\usepackage{graphics}

\usepackage{amsmath} 
\usepackage{amsthm} 
\usepackage{url} 
\usepackage{stmaryrd}
\usepackage{tikz}
\usepackage{physics}
\usepackage{bm}
\usepackage{booktabs}
\usepackage{subcaption}

\numberwithin{equation}{section}

\allowdisplaybreaks

\begin{document}

\title{Direct Monte Carlo Computation of the 't~Hooft Partition Function}


\author[1]{Okuto Morikawa}
\affil[1]{Interdisciplinary Theoretical and Mathematical Sciences Program
(iTHEMS), RIKEN, Wako 351-0198, Japan}

\author[2]{Hiroshi Suzuki}
\affil[2]{Department of Physics, Kyushu University, 744 Motooka, Nishi-ku,
Fukuoka 819-0395, Japan}





\begin{abstract}%
The 't~Hooft partition function~$\mathcal{Z}_{\text{tH}}[E;B]$ of an $SU(N)$ gauge
theory with the $\mathbb{Z}_N$ 1-form symmetry is defined as the Fourier
transform of the partition function~$\mathcal{Z}[B]$ with respect to the
spatial-temporal components of the 't~Hooft flux~$B$. Its large volume
behavior detects the quantum phase of the system. When the integrand of the
functional integral is real-positive, the latter partition
function~$\mathcal{Z}[B]$ can be numerically computed by a Monte Carlo
simulation of the $SU(N)/\mathbb{Z}_N$ gauge theory, just by counting the
number of configurations of a specific 't~Hooft flux~$B$. We carry out this
program for the $SU(2)$ pure Yang--Mills theory with the vanishing
$\theta$-angle by employing a newly-developed hybrid Monte Carlo (HMC)
algorithm (the halfway HMC) for the $SU(N)/\mathbb{Z}_N$ gauge theory. The
numerical result clearly shows that all non-electric fluxes are ``light'' as
expected in the ordinary confining phase with the monopole condensate.
Invoking the Witten effect on~$\mathcal{Z}_{\text{tH}}[E;B]$, this also indicates
the oblique confinement at~$\theta=2\pi$ with the dyon condensate.
\end{abstract}

\subjectindex{B01,B03,B31}

\maketitle

\section{Introduction}
\label{sec:1}
For an $SU(N)$ gauge theory with the $\mathbb{Z}_N$ 1-form
symmetry~\cite{Gaiotto:2014kfa}, such as the pure Yang--Mills theory or
the $\mathcal{N}=1$ and~$\mathcal{N}=1^*$ supersymmetric Yang--Mills
theories~\cite{Donagi:1995cf}, one can introduce the 't~Hooft
flux~$B:=(B_{12},B_{13},B_{14},B_{23},B_{24},B_{34})\in\mathbb{Z}_N$ by the twisted
boundary conditions on the 4~torus~$T^4$~\cite{tHooft:1979rtg,tHooft:1981sps}.
The 't~Hooft partition function~$\mathcal{Z}_{\text{tH}}[E;B]$ is then defined as
the Fourier transform of the partition function~$\mathcal{Z}[B]$ with respect
to the spatial-temporal components of the 't~Hooft
flux~$B$~\cite{tHooft:1979rtg,tHooft:1981sps}:
\begin{equation}
   \mathcal{Z}_{\text{tH}}[E_1,E_2,E_3;B_{12},B_{23},B_{31}]
   :=\frac{1}{N^3}\sum_{B_{14},B_{24},B_{34}=0}^{N-1}
   \exp\left(\frac{2\pi i}{N}\sum_{i=1}^3E_iB_{i4}\right)
   \mathcal{Z}[B].
\label{eq:(1.1)}
\end{equation}
On the left-hand side, $E_i$ and~$B_{ij}$ are referred to as the electric and
magnetic fluxes, respectively. The large volume behavior
of~$\mathcal{Z}_{\text{tH}}[E;B]$ detects the quantum phase (i.e., confinement,
Higgs, or Coulomb) of the system~\cite{tHooft:1979rtg,tHooft:1981sps}. See
also~Ref.~\cite{Tomboulis:1985ah}. Recently, consideration of the 't~Hooft
partition function~$\mathcal{Z}_{\text{tH}}[E;B]$ has been
revived~\cite{Kitano:2017jng,Tanizaki:2022ngt,Nguyen:2023fun}, largely
motivated by the perspective of the generalized
symmetries~\cite{Gaiotto:2014kfa}, in particular in connection with the study
in~Ref.~\cite{Gaiotto:2017yup}; see also Refs.~\cite{Yamazaki:2017ulc,Hayashi:2023wwi,Hayashi:2024qkm,Hayashi:2024yjc,Hayashi:2024psa} for recent related
studies.

Now, when the integrand of the functional integral is real-positive as the case
in the pure Yang--Mills theory with the vanishing $\theta$-angle, the partition
function~$\mathcal{Z}[B]$ on the right-hand side of~Eq.~\eqref{eq:(1.1)}, or
more precisely the ratio~$\mathcal{Z}[B]/\mathcal{Z}[0]$, may be numerically
computed by a Monte Carlo simulation~\cite{Kovacs:2000sy,deForcrand:2001nd}.
Traditionally, this ratio is computed by ``reweighting'' the difference of
lattice actions with and without~$B$~\cite{Kovacs:2000sy,deForcrand:2001nd}. In
the present paper, we employ a Monte Carlo simulation of the
$SU(N)/\mathbb{Z}_N$ Yang--Mills theory on the basis of a recently developed
hybrid Monte Carlo (HMC) algorithm (the halfway HMC)~\cite{Abe:2024fpt} in
which the 't~Hooft flux~$B$ (the total flux of the $\mathbb{Z}_N$ 2-form flat
gauge field) is explicitly treated as a dynamical variables.\footnote{For a
more ``traditional'' approach to the $SU(N)/\mathbb{Z}_N$ Yang--Mills theory
using the plaquette action in the adjoint representation,
see~Refs.~\cite{Halliday:1981te,Creutz:1982ga}. See also
Refs.~\cite{Edwards:1998dj,deForcrand:2002vs,Halliday:1981tm}.} In this way,
each configuration generated in the Monte Carlo simulation possesses a various
but definite value of the 't~Hooft flux~$B$. Then, just by counting the number
of configurations of a specific~$B$, we can obtain the partition
function~$\mathcal{Z}[B]$.\footnote{%
We expect that our computational method is free from the overlap problem that
requires an elaborate computational trick in the reweighting
approach~\cite{Kovacs:2000sy,deForcrand:2001nd}.} The 't~Hooft partition
function~$\mathcal{Z}_{\text{tH}}[E;B]$ is then given by~Eq.~\eqref{eq:(1.1)}.
Explicitly, we carry out this program for the $SU(2)$ pure Yang--Mills theory
with a vanishing $\theta$-angle.\footnote{In this paper, we study the
partition function in the presence of the 't~Hooft flux, not the 't~Hooft
line~\cite{tHooft:1977nqb}. The former is the total flux of the $\mathbb{Z}_N$
2-form \emph{flat\/} gauge field such that~$\mathrm{d}B=0\bmod N$ (i.e.,
elements of~$H^2(T^4,\mathbb{Z}_N)$), while for the latter $\mathrm{d}B$ is
given by the Poincar\'e dual of the line~\cite{Aharony:2013hda}.
Computationally, the study of the
latter~\cite{Hoelbling:1998gv,Hoelbling:1999gr,Hoelbling:2000su,deForcrand:2000fi,DelDebbio:2000cb} is more demanding.} Our numerical result below clearly
shows that all nonelectric fluxes are ``light'' as expected in the ordinary
confining phase with the monopole
condensate~\cite{tHooft:1979rtg,tHooft:1981sps}. Although the study of the
't~Hooft partition function has a long history, to our knowledge, this is the
first attempt to measure the 't~Hooft partition function with all possible
combinations of the 't~Hooft flux by a lattice Monte Carlo simulation.

As in our numerical calculation, when all cycles of the 4~torus~$T^4$ possess
an equal radius~$L$, the partition function~$\mathcal{Z}[B]$ enjoys the
Euclidean $90^\circ$ rotational invariance (see Appendix~\ref{sec:A}) and, as
the consequence of this, $\mathcal{Z}_{\text{tH}}[E;B]$ obeys the duality
equation~\cite{tHooft:1979rtg,tHooft:1981sps},
\begin{align}
   &\mathcal{Z}_{\text{tH}}[E_1,E_2,E_3;B_{12},B_{23},B_{31}]
\notag\\
   &=\frac{1}{N^2}\sum_{B_{23}',B_{31}',E_1',E_2'=0}^{N-1}
   \exp\left[\frac{2\pi i}{N}
   \left(E_1B_{23}'+E_2B_{31}'-B_{23}E_1'-B_{31}E_2'\right)\right]
\notag\\
   &\qquad\qquad\qquad\qquad\qquad{}
   \times{}\mathcal{Z}_{\text{tH}}[E_1',E_2',E_3;B_{12},B_{23}',B_{31}'].
\label{eq:(1.2)}
\end{align}
We observe that to fairly good numerical accuracy our numerical result
for~$\mathcal{Z}_{\text{tH}}[E;B]$ fulfills this equation, providing a consistency
check of the computation. We may even take the average of~$\mathcal{Z}[B]$ over
Euclidean $90^\circ$ rotations so that $\mathcal{Z}_{\text{tH}}[E;B]$ automatically
fulfills this duality equation within the numerical error (see below).

\section{Direct computation of the 't~Hooft partition function}
\label{sec:2}
Our lattice action on a periodic lattice of size~$L$,
$\Gamma:=(\mathbb{Z}/L\mathbb{Z})^4$, for the $SU(N)/\mathbb{Z}_N$ theory is
given by~\cite{Mack:1978kr,Ukawa:1979yv,Srednicki:1980gb,Seiler:1982pw}
(see also Ref.~\cite{Kapustin:2014gua}):
\begin{equation}
   S:=-\beta\sum_{x\in\Gamma}\sum_{\mu<\nu}
   \frac{1}{N}\Re\tr\left[e^{-2\pi iB_{\mu\nu}(x)/N}P(x,\mu,\nu)-\bm{1}\right],
\label{eq:(2.1)}
\end{equation}
where $\beta$ is the bare coupling and the plaquette variables~$P(x,\mu,\nu)$
are given by
\begin{equation}
   P(x,\mu,\nu):=U(x,\mu)U(x+\Hat{\mu},\nu)
   U(x+\Hat{\nu},\mu)^\dagger U(x,\nu)^\dagger
\label{eq:(2.2)}
\end{equation}
from $SU(N)$ link variables. The integer field~$B_{\mu\nu}(x)$
in~Eq.~\eqref{eq:(2.1)} is given by the 't~Hooft flux by
\begin{equation}
   B_{\mu\nu}(x)=\begin{cases}
   B_{\mu\nu}&\text{for $x_\mu=L-1$ and~$x_\nu=L-1$},\\
   0&\text{otherwise}\\
   \end{cases}
\label{eq:(2.3)}
\end{equation}
with $B_{\mu\nu}=\{0,1,\dotsc,N-1\}\bmod N$ (we set~$B_{\nu\mu}=-B_{\mu\nu}$). For
the Boltzmann weight~$e^{-S}$, we generated configurations of~$(U,B)$ for~$N=2$
by employing the halfway HMC. We refer the reader to~Ref.~\cite{Abe:2024fpt}
for details of our numerical simulation.\footnote{Our numerical codes can be
found in \url{https://github.com/o-morikawa/Gaugefields.jl}, which is based on
\texttt{Gaugefields.jl} in the JuliaQCD package~\cite{Nagai:2024yaf}.} The
partition function~$\mathcal{Z}[B]$ with a particular 't~Hooft flux, say,
$(B_{12},B_{13},B_{14},B_{23},B_{24},B_{34})=(0,0,0,1,1,1)$ can be obtained as the
expectation value of the operator,
\begin{equation}
   \mathcal{O}_{(0,0,0,1,1,1)}:=\frac{1}{2^6}\delta_{B,(0,0,0,1,1,1)}.
\label{eq:(2.4)}
\end{equation}
The expectation value, however, can be computed by simply counting the number
of configurations with the flux~$(0,0,0,1,1,1)$ and dividing it by the total
number of configurations.\footnote{We store gauge field configurations
generated by the halfway HMC~\cite{Abe:2024fpt} under filenames such as
\texttt{U\_beta2.6\_L20\_F111001\_7020.txt}, where the number \texttt{111001}
represents the value of the flux~$(B_{12},B_{13},B_{14},B_{23},B_{24},B_{34})$ of
that configuration.}

We may further consider the average of~$\mathbb{Z}[B]$ over the Euclidean
$90^\circ$ orthogonal rotations. See Appendix~\ref{sec:A} for a list of the
irreducible representations. For instance, corresponding to the dimension~4
irreducible representation in~Eq.~\eqref{eq:(A8)}, Eq.~\eqref{eq:(2.4)} may
be replaced by
\begin{equation}
   \Bar{\mathcal{O}}_{(0,0,0,1,1,1)}
   :=\frac{1}{2^6}\frac{1}{4}
   \left[\delta_{B,(0,0,0,1,1,1)}+\delta_{B,(0,1,1,0,0,1)}
   +\delta_{B,(1,0,1,0,1,0)}+\delta_{B,(1,1,0,1,0,0)}\right].
\label{eq:(2.5)}
\end{equation}
We will see that this prescription reduces the statistical error considerably.

In what follows, we show the results using 2590 configurations for $\beta=2.6$
and~$L=20$. No attempt to find the continuum limit is made because the
behavior is almost the same for all lattice parameters considered
in~Ref.~\cite{Abe:2024fpt}.

\begin{figure}[htbp]
\centering
\includegraphics[width=12cm]{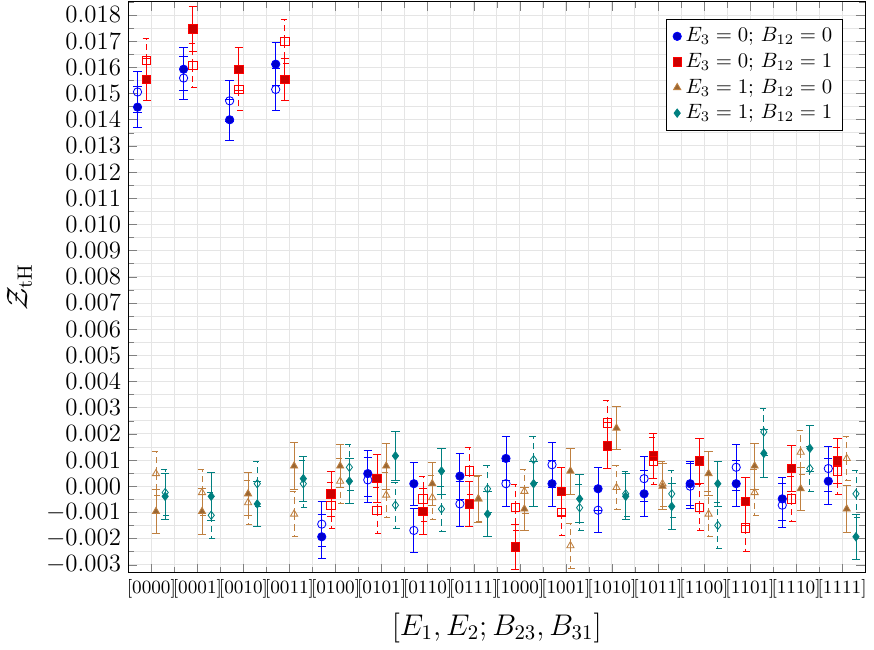}
\caption{The 't~Hooft partition function~$\mathcal{Z}_{\text{tH}}[E;B]$ for all
possible combinations of the 't~Hooft fluxes, $E_i$ and~$B_{ij}$. $\beta=2.6$
and~$L=20$. The statistical errors are estimated by the jackknife method. No
average over Euclidean $90^\circ$ rotations is taken. The filled symbols
represent $\mathcal{Z}_{\text{tH}}[E;B]$ computed from~$\mathcal{Z}[B]$
by~Eq.~\eqref{eq:(1.1)} and the unfilled symbols represent the right-hand side
of the duality equation~\eqref{eq:(1.2)}. We observe that the duality equation
holds to a fairly good accuracy.}
\label{fig:1}
\end{figure}

First, in~Fig.~\ref{fig:1}, we plot the 't~Hooft partition
function~$\mathcal{Z}_{\text{tH}}[E;B]$ for all possible combinations of 't~Hooft
fluxes, $E_i$ and~$B_{ij}$. The statistical errors are estimated by the
jackknife method.\footnote{We found that statistical errors are almost
saturated at the bin size~$\sim1$. We think that this is a consequence of a
shortness of the autocorrelation length in the present HMC
algorithm~\cite{Abe:2024fpt}.} In this figure, no average over Euclidean
$90^\circ$ rotations is taken. The filled symbols represent
$\mathcal{Z}_{\text{tH}}[E;B]$ computed from~$\mathcal{Z}[B]$
by~Eq.~\eqref{eq:(1.1)} and the unfilled symbols represent the right-hand side
of the duality equation~\eqref{eq:(1.2)}. We thus observe that the duality
equation~\eqref{eq:(1.2)} holds to a fairly good accuracy.
In Ref.~\cite{Nguyen:2023fun}, the authors showed that the 't~Hooft partition
function~$\mathcal{Z}_{\text{tH}}[E;B]$ is real semi-positive as far as the
reflection positivity holds. The plots in~Fig.~\ref{fig:1} are also consistent
with this property within the errors.

From~Fig.~\ref{fig:1}, we immediately observe that the 't~Hooft partition
functions~$\mathcal{Z}_{\text{tH}}[E;B]$ are clearly classified into two classes
depending on the value of the flux; one gives
$\mathcal{Z}_{\text{tH}}[E;B]/\mathcal{Z}_{\text{tH}}[E=0;B=0]\sim1$ and another
gives $\mathcal{Z}_{\text{tH}}[E;B]/\mathcal{Z}_{\text{tH}}[E=0;B=0]\sim0$. The
former class of fluxes is called ``light'' and the latter class is called
``heavy''~\cite{tHooft:1979rtg,tHooft:1981sps}. From the duality
equation~\eqref{eq:(1.2)}, it can be argued that, when $E_3$ and~$B_{12}$ are
fixed, there are only $0$ or~$N^2=4$ combinations of light fluxes among
totally $N^4=16$ combinations of fluxes~\cite{tHooft:1979rtg,tHooft:1981sps}.
See also Ref.~\cite{Nguyen:2023fun}. We clearly see that this assertion holds
in~Fig.~\ref{fig:1}. We also observe that all fluxes with $E_i\neq0$ are heavy.
In other words, no light fluxes possess electric flux, $E_i=0$. This is
expected in the ordinary confining phase in which the magnetic monopole
condensates~\cite{tHooft:1979rtg,tHooft:1981sps}.%
\footnote{The inverse Fourier transform of~$\mathcal{Z}_{\text{tH}}[E;B]$ with
this behavior in the large volume limit gives
$\mathcal{Z}[B]/\mathcal{Z}[B=0]\sim1$ for any~$B$. In modern
language~\cite{Gaiotto:2014kfa,Nguyen:2023fun}, this implies that one can
define the $\mathbb{Z}_N$ 1-form symmetry in the low-energy theory, whose
symmetry operator is spanned by a 2-surface given by the Poincar\'e dual of a
given flux~$B$. When $B_{34}=1$ and other components vanish, the symmetry
operator is spanned by a 2-surface in the $12$-direction. Then a move of a
fundamental Wilson line extending in the $4$-direction along the cycle in the
$3$-direction produces the factor~$e^{\pm2\pi i/N}$. This factor forbids the
expectation value of the Wilson line (or the Polyakov line) and implies that
the $\mathbb{Z}_N$ 1-form symmetry is not spontaneously broken; this is a
characterization of the ordinary confining phase.}

\begin{figure}[htbp]
\centering
\includegraphics[width=12cm]{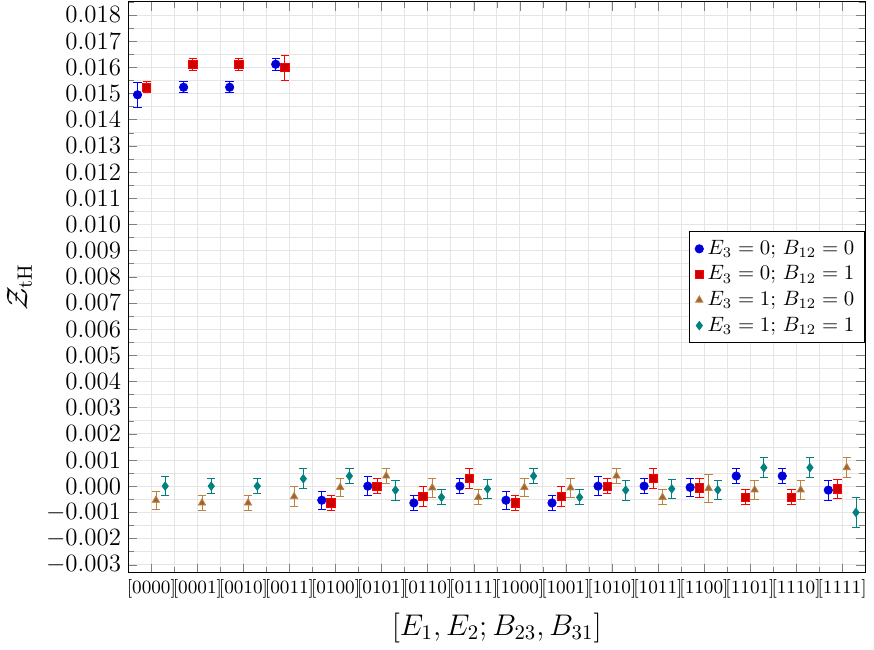}
\caption{The 't~Hooft partition function~$\mathcal{Z}_{\text{tH}}[E;B]$ for all
possible combinations of 't~Hooft fluxes, $E_i$ and~$B_{ij}$. $\beta=2.6$
and~$L=20$. The errors are estimated by the jackknife method. The average
of~$\mathcal{Z}[B]$ over Euclidean $90^\circ$ rotations is made.}
\label{fig:2}
\end{figure}

Figure~\ref{fig:2} is the same as~Fig.~\ref{fig:1}, but the average over the
Euclidean $90^\circ$ rotations is taken in~$\mathcal{Z}[B]$. The statistical
errors become smaller as anticipated; the duality equation is automatically
fulfilled and the separation into light fluxes and heavy fluxes becomes
clearer.

Our result thus illustrates that a direct numerical computation of the 't~Hooft
partition function can become a useful approach to study the quantum phase of
$SU(N)$ gauge theories with the $\mathbb{Z}_N$ 1-form symmetry, at least as far
as the integrand of the functional integral is real-positive; when the complex
phase of the integrand is small, it could be treated by reweighting.

The behavior $\mathcal{Z}_{\text{tH}}[E;B]/\mathcal{Z}_{\text{tH}}[E=0;B=0]\to1$ or
$\mathcal{Z}_{\text{tH}}[E;B]/\mathcal{Z}_{\text{tH}}[E=0;B=0]\to0$ is an asymptotic
one in the large volume limit~\cite{tHooft:1979rtg,tHooft:1981sps} and the way
of approaching it provides useful information on the (dual) string tension.
Unfortunately, we found that it is difficult at the moment to determine the
sting tension in our simulation with the present lattice parameters and
statistics.

\begin{figure}[htbp]
\centering
\includegraphics[width=12cm]{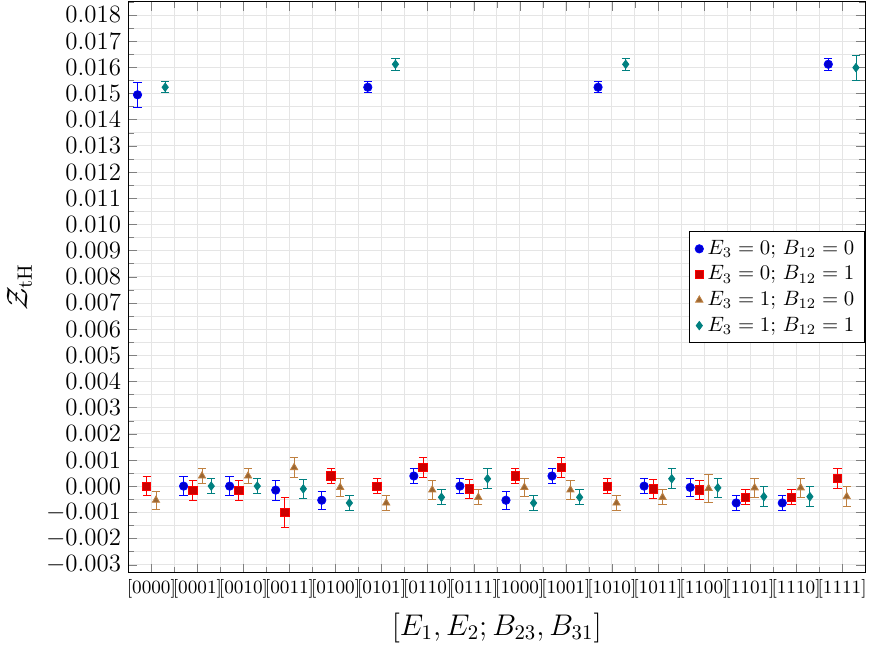}
\caption{The 't~Hooft partition function with~$\theta=2\pi$,
$\mathcal{Z}_{\text{tH},\theta=2\pi}[E;B]$ for all possible combinations of 't~Hooft
fluxes, $E_i$ and~$B_{ij}$. $\beta=2.6$ and~$L=20$. This plot was obtained by
simply applying the Witten effect~\eqref{eq:(2.7)} with~$\theta=0$ to data
in~Fig.~\ref{fig:2}.}
\label{fig:3}
\end{figure}

We may generalize the partition function as~$\mathcal{Z}_\theta[B]$ by
multiplying the $\theta$-term $e^{-i\theta Q}$ to the Boltzmann weight. In the
pure $SU(N)$ Yang--Mills theory, $Q\in\mathbb{Z}$, but in the presence of the
't~Hooft flux, $Q$ becomes fractional as~\cite{tHooft:1981nnx,vanBaal:1982ag}:
\begin{equation}
   Q=-\frac{1}{N}\frac{\varepsilon_{\mu\nu\rho\sigma}B_{\mu\nu}B_{\rho\sigma}}{8}
   +\mathbb{Z}.
\label{eq:(2.6)}
\end{equation}
It is possible to construct a geometric definition of~$Q$ on the lattice that
possesses this property~\cite{Abe:2023ncy} by requiring the $\mathbb{Z}_N$
1-form gauge symmetry in the construction in~Ref.~\cite{Luscher:1981zq}.
Noting Eq.~\eqref{eq:(2.6)}, under the shift $\theta\to\theta+2\pi$, one finds
that the corresponding 't~Hooft partition
function~$\mathcal{Z}_{\text{tH},\theta}[E;B]$~\eqref{eq:(1.1)} behaves as
\begin{equation}
   \mathcal{Z}_{\text{tH},\theta+2\pi}[E_1,E_2,E_3;B_{12},B_{23},B_{31}]
   =\mathcal{Z}_{\text{tH},\theta}[E_1+B_{23},E_2+B_{31},E_3+B_{12};B_{12},B_{23},B_{31}].
\label{eq:(2.7)}
\end{equation}
This is the Witten effect~\cite{Witten:1979ey} on the 't~Hooft partition
function. Invoking this relation with~$\theta=0$, from~Figs.~\ref{fig:1}
and~\ref{fig:2}, we infer that fluxes
with~$E_1+B_{23}=E_2+B_{31}=E_3+B_{12}=0\bmod2$ are light for~$\theta=2\pi$;
see Fig.~\ref{fig:3} for the 't~Hooft partition
function~$\mathcal{Z}_{\text{tH},\theta=2\pi}[E;B]$. This pattern of light fluxes is
an indication of the oblique confinement, in which the dyons
condensate~\cite{tHooft:1979rtg,tHooft:1981bkw,tHooft:1981sps}.

\section{Conclusion}
\label{sec:3}
The pattern exhibited by the 't~Hooft partition
function~$\mathcal{Z}_{\text{tH}}[E;B]$ as the function of the flux, such as that
in Figs.~\ref{fig:2} and~\ref{fig:3}, clearly indicates the quantum phase of
the system~\cite{tHooft:1979rtg,tHooft:1981sps}. In this paper, by employing
the halfway HMC~\cite{Abe:2024fpt} for the simplest example, the $SU(2)$ pure
Yang--Mills theory, we illustrated that a direct Monte Carlo calculation of the
pattern is feasible, at least as long as the integrand of the functional
integral is real-positive; when the complex phase of the integrand is small, it
could be treated by reweighting. Our methodology itself is quite general so we
hope to carry out analyses in more intriguing situations: It should be
interesting to study finite temperature cases (such as
in~Ref.~\cite{deForcrand:2001nd}) to observe the confining/deconfining phase
transition. The inclusion of matter fields such as the adjoint Higgs and
adjoint fermions will give a more intricate pattern of the 't~Hooft partition
function. We hope to tackle these problems in the near future.

\section*{Acknowledgments}
We would like to thank Motokazu Abe for discussions spanning a long period of
time, which finally led to this work. We also thank Philippe de Forcrand, Yui
Hayashi, Yuki Nagai, Soma Onoda, Yuya Tanizaki, Akio Tomiya, and Hiromasa
Watanabe for helpful discussions.
We appreciate the opportunity of discussion during the YITP--RIKEN iTHEMS
conference ``Generalized symmetries in QFT 2024'' (YITP-W-24-15) in the
execution of this work.
The numerical computations in this paper were carried out on Genkai, a
supercomputer system of the Research Institute for Information Technology
(RIIT), Kyushu University.
O.M.\ acknowledges the RIKEN Special Postdoctoral Researcher Program.
The work of H.S. was partially supported by a Japan Society for the Promotion
of Science (JSPS) Grant-in-Aid for Scientific Research, JP23K03418.

\appendix

\section{Euclidean $90^\circ$ discrete rotations on the $\mathbb{Z}_2$ flux in
4D and 3D}
\label{sec:A}
On the components of a second rank antisymmetric tensor,
\begin{equation}
   B_1:=B_{12},\quad
   B_2:=B_{13},\quad
   B_3:=B_{14},\quad
   B_4:=B_{23},\quad
   B_5:=B_{24},\quad
   B_6:=B_{34},
\label{eq:(A1)}
\end{equation}
Euclidean $90^\circ$ rotations on the $\mu\nu$-plane are represented as
$B_a\to\sum_{b=1}^6(\Lambda_{\mu\nu})_{ab}B_b$, where the representation matrices
are given by,
\begin{align}
    &\Lambda_{12}=\begin{pmatrix}
        1& 0& 0& 0& 0& 0\\
        0& 0& 0& -1& 0& 0\\
        0& 0& 0& 0& -1& 0\\
        0& 1& 0& 0& 0& 0\\
        0& 0& 1& 0& 0& 0\\
        0& 0& 0& 0& 0& 1\\
    \end{pmatrix},
\qquad
    \Lambda_{13}=\begin{pmatrix}
        0& 0& 0& 1& 0& 0\\
        0& 1& 0& 0& 0& 0\\
        0& 0& 0& 0& 0& -1\\
        -1& 0& 0& 0& 0& 0\\
        0& 0& 0& 0& 1& 0\\
        0& 0& 1& 0& 0& 0\\
    \end{pmatrix},
\notag\\
    &\Lambda_{14}=\begin{pmatrix}
        0& 0& 0& 0& 1& 0\\
        0& 0& 0& 0& 0& 1\\
        0& 0& 1& 0& 0& 0\\
        0& 0& 0& 1& 0& 0\\
        -1& 0& 0& 0& 0& 0\\
        0& -1& 0& 0& 0& 0\\
    \end{pmatrix},
\qquad
    \Lambda_{23}=\begin{pmatrix}
        0& -1& 0& 0& 0& 0\\
        1& 0& 0& 0& 0& 0\\
        0& 0& 1& 0& 0& 0\\
        0& 0& 0& 1& 0& 0\\
        0& 0& 0& 0& 0& -1\\
        0& 0& 0& 0& 1& 0\\
    \end{pmatrix},
\notag\\
    &\Lambda_{24}=\begin{pmatrix}
        0& 0& -1& 0& 0& 0\\
        0& 1& 0& 0& 0& 0\\
        1& 0& 0& 0& 0& 0\\
        0& 0& 0& 0& 0& 1\\
        0& 0& 0& 0& 1& 0\\
        0& 0& 0& -1& 0& 0\\
    \end{pmatrix},
\qquad
    \Lambda_{34}=\begin{pmatrix}
        1& 0& 0& 0& 0& 0\\
        0& 0& -1& 0& 0& 0\\
        0& 1& 0& 0& 0& 0\\
        0& 0& 0& 0& -1& 0\\
        0& 0& 0& 1& 0& 0\\
        0& 0& 0& 0& 0& 1\\
    \end{pmatrix}.
\label{eq:(A2)}
\end{align}
Some configurations of the 't~Hooft flux are thus related by these
transformations. For~$N=2$, for which $B_a=\{0,1\}\bmod2$, the configurations
are classified into the following 11 irreducible representations,
Eqs.~\eqref{eq:(A3)}--\eqref{eq:(A9)} and combinations obtained by the
interchange~$0\leftrightarrow1$ in~Eqs.~\eqref{eq:(A3)}--\eqref{eq:(A6)}:
\begin{equation}
(B_1,B_2,B_3,B_4,B_5,B_6)=
(0,0,0,0,0,0),
\label{eq:(A3)}
\end{equation}
\begin{subequations}
\begin{align}
&(0,0,0,0,0,1),\quad
(0,0,0,0,1,0),\quad
(0,0,1,0,0,0),
\label{eq:(A4a)}\\
&(0,0,0,1,0,0),\quad
(0,1,0,0,0,0),\quad
(1,0,0,0,0,0),
\label{eq:(A4b)}
\end{align}
\label{eq:(A4)}
\end{subequations}
\begin{subequations}
\begin{align}
&(0,0,0,0,1,1),\quad
(0,0,1,0,0,1),\quad
(0,0,1,0,1,0),
\label{eq:(A5a)}\\
&(0,1,0,1,0,0),\quad
(1,0,0,1,0,0),\quad
(1,1,0,0,0,0),
\label{eq:(A5b)}\\
&(0,0,0,1,0,1),\quad
(0,0,0,1,1,0),\quad
(0,1,0,0,0,1),
\notag\\
&(0,1,1,0,0,0),\quad
(1,0,0,0,1,0),\quad
(1,0,1,0,0,0),
\label{eq:(A5c)}
\end{align}
\label{eq:(A5)}
\end{subequations}
\begin{align}
(0,0,1,1,0,0),\quad
(0,1,0,0,1,0),\quad
(1,0,0,0,0,1),
\label{eq:(A6)}
\end{align}
\begin{subequations}
\begin{gather}
(0,0,1,0,1,1),
\label{eq:(A7a)}\\
(0,1,0,1,0,1),\quad
(1,0,0,1,1,0),\quad
(1,1,1,0,0,0),
\label{eq:(A7b)}
\end{gather}
\label{eq:(A7)}
\end{subequations}
\begin{subequations}
\begin{gather}
(1,1,0,1,0,0),
\label{eq:(A8a)}\\
(0,0,0,1,1,1),\quad
(0,1,1,0,0,1),\quad
(1,0,1,0,1,0),\quad
\label{eq:(A8b)}
\end{gather}
\label{eq:(A8)}
\end{subequations}
\begin{subequations}
\begin{align}
&(0,0,1,1,0,1),\quad
(0,0,1,1,1,0),\quad
(0,1,0,0,1,1),
\notag\\
&(0,1,1,0,1,0),\quad
(1,0,0,0,1,1),\quad
(1,0,1,0,0,1),
\label{eq:(A9a)}\\
&(0,1,0,1,1,0),\quad
(0,1,1,1,0,0),\quad
(1,0,0,1,0,1),
\notag\\
&(1,0,1,1,0,0),\quad
(1,1,0,0,0,1),\quad
(1,1,0,0,1,0).
\label{eq:(A9b)}
\end{align}
\label{eq:(A9)}
\end{subequations}

We may thus take the average of the partition function~$\mathcal{Z}[B]$ over
elements within each of these irreducible representations. This average is
taken in~Figs.~\ref{fig:2} and~\ref{fig:3}.

For finite temperatures, $\mathcal{Z}[B]$ is invariant only under 3D spatial
$90^\circ$ rotations generated by~$\Lambda_{12}$, $\Lambda_{13}$,
and~$\Lambda_{23}$. Then each irreducible representation is decomposed into
smaller multiplets labeled by a roman index in the equation number, such as
``a'' in~\eqref{eq:(A5a)}. We may then take the average of~$\mathcal{Z}[B]$
over elements within these smaller multiplets.



%



\let\doi\relax










\end{document}